\documentclass[runningheads]{llncs}

\usepackage[T1]{fontenc}
\usepackage{amsmath,amssymb}
\usepackage{booktabs}
\usepackage{algorithmic,algorithm,float}
\usepackage{graphicx}
\usepackage{adjustbox}
\usepackage{tikz}
\usetikzlibrary{arrows.meta,decorations.pathmorphing}
\usepackage{float}
\usepackage{enumitem}

\usepackage{pmboxdraw}

\makeatletter
\renewcommand{\algorithmiccomment}[1]{\hfill{\color{blue!90!black}//~#1}}
\makeatother

\usepackage{hyperref}
\usepackage[capitalise]{cleveref}

\usepackage[normalem]{ulem}

\newcommand{\favoritelabelsep}{4pt}
\newcommand{\favoriteitemsep}{2pt}

\begin{document}

\title{Fast Rational Univariate Representation via Gaussian Elimination}
\titlerunning{Fast RUR via Gaussian Elimination}

\author{Alexander Demin\inst{1}\thanks{
Alexander Demin has been supported by an ERC-2023-ADG grant for the ODELIX project (number 101142171).
\\[0.6em]
\noindent\begin{minipage}{0.72\linewidth}
{\it Funded by the European Union. Views and opinions expressed are however those of the author(s) only and do not necessarily reflect those of the European Union or the European Research Council Executive Agency. Neither the European Union nor the granting authority can be held responsible for them.}
\end{minipage}
\hfill
\begin{minipage}{0.24\linewidth}
\centering
\protect\includegraphics[height=1.1cm]{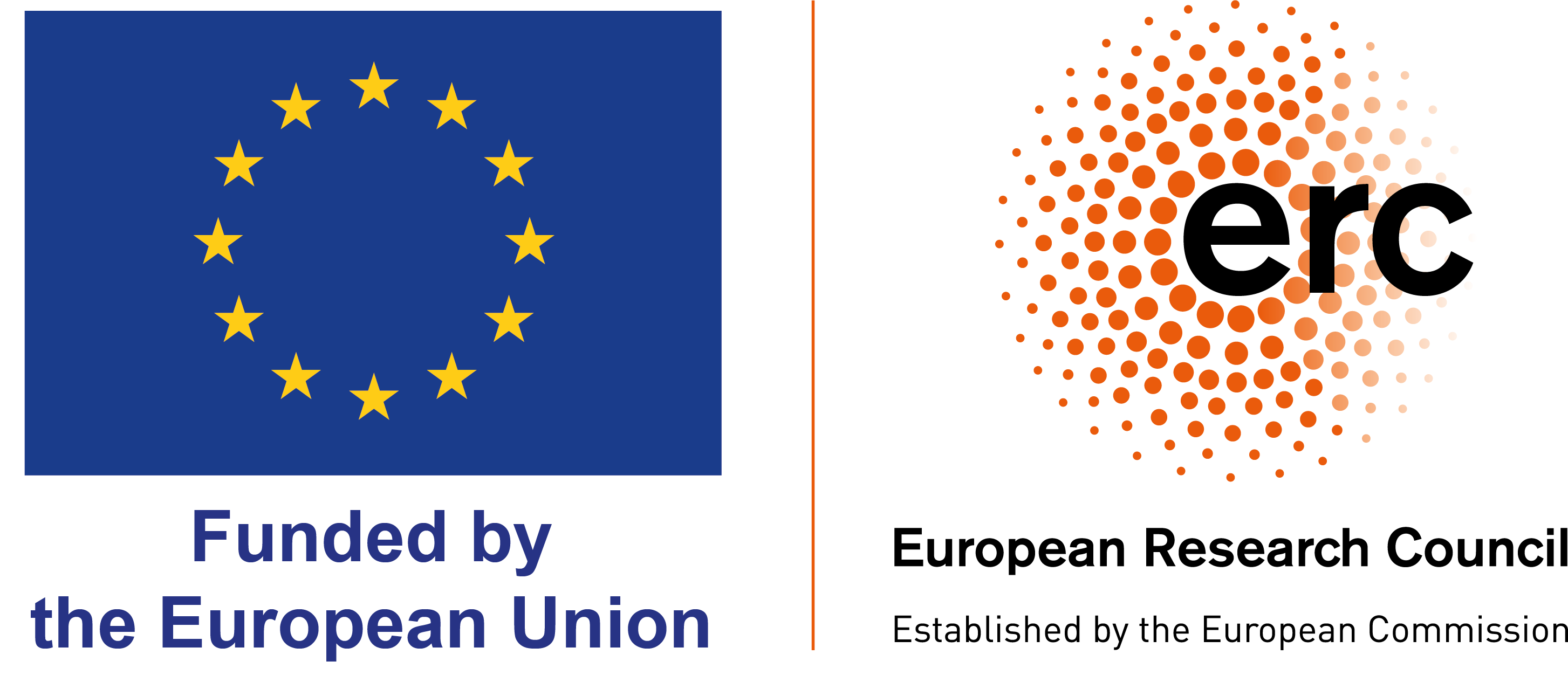}
\end{minipage}
} \and Fabrice Rouillier\inst{2}}
\authorrunning{Demin and Rouillier}

\institute{
Laboratoire d'informatique de l'{\'E}cole polytechnique,\\LIX, UMR 7161, CNRS,\\1 rue Honor{\'e} d'Estienne d'Orves,\\91120 Palaiseau, France\\
\email{demin@lix.polytechnique.fr}
\and
Sorbonne Universit\'e, CNRS, Inria, Paris, France \\
\email{Fabrice.Rouillier@inria.fr}
}

\maketitle

\begin{abstract}
In this note, we present \href{https://newrur.gitlabpages.inria.fr/RationalUnivariateRepresentation.jl/}{RationalUnivariateRepresentation.jl}, a Julia package for computing rational univariate representations of zero-dimensional polynomial systems.
The package uses dense linear algebra and Gaussian elimination for the FGLM-like stage.
The purpose of this contribution is to advocate for this choice and explain the implementation details that turn the algorithm into practical software.
In particular, we show that our implementation can compute guaranteedly correct parametrizations of ideals with thousands of solutions within seconds.

\end{abstract}

\section{Introduction}

Let $k$ be a field and $\bar{k}$ its algebraic closure. Let $I \subseteq k[x_1,\ldots,x_n]$ be a zero-dimensional ideal. A rational univariate representation~\cite{rouillier99} of the roots of $I$ in $\bar{k}^n$ consists of a linear form $t = a_1 x_1 + \cdots + a_n x_n$ that is injective on the roots, where $a_1,\ldots,a_n \in k$, together with polynomials $f, f_1, \dots, f_n \in k[T]$. Each root of $I$ can be recovered from a root $\beta \in \bar{k}$ of $f$ via $x_i = f_i(\beta)/f'(\beta)$. Our goal is to compute such parametrizations of the roots of the radical $\sqrt{I}$.

When the ideal is in shape position, 
a lexicographic Gr{\"o}bner basis of $I$ yields such a parametrization, and it can be efficiently computed from a degrevlex Gr{\"o}bner basis using the FGLM algorithm~\cite{fglm}. In general, with a random change of variables, the radical can be placed in shape position with high probability. Several strategies exploit the sparsity of the multiplication matrices arising in FGLM through the Wiedemann algorithm~\cite{Wiedemann} or Block Krylov methods~\cite{faugere-mou,hnrs-sparse-fglm}; these are implemented in solvers such as msolve~\cite{msolve} or Giac~\cite{parisse-rur}.

In this paper, we make several practical observations. For example, the multiplication matrices arising in the FGLM algorithm are often only moderately sparse (see \Cref{sec:on-sparsity}), which makes sparse linear algebra less compelling than one might expect. We then discuss how these observations motivate the use of classical Gaussian elimination for computing parametrizations.

We present an optimized deterministic implementation of the algorithm of Demin, Rouillier, and Ruiz~\cite{demin-rouillier-ruiz}, based on dense linear algebra and classical Gaussian elimination, and show experimentally that this design leads to competitive performance with state-of-the-art software.

\section{Background}

\subsection{Shape position}\label{sec:-2-1}

Let $I \subseteq k[x_1,\ldots,x_n]$ be a zero-dimensional ideal.
Let $V(I) \subseteq \bar{k}^n$ denote the algebraic variety of $I$.
Let $D$ be the dimension of the quotient algebra $k[x_1,\ldots,x_n] / I$ as a $k$-vector space. Equivalently, $D$ is the number of roots of $I$ counted with multiplicities.

A particularly simple parametrization arises when the ideal is in shape position:

\begin{definition}[Shape position]
    An ideal $I$ is said to be in shape position with respect to the indeterminate $x_n$ if the reduced Gr{\"o}bner basis with respect to the lexicographical ordering with $x_n < \ldots < x_1$ is
    \begin{equation*}\label{eq:shape}
        G = \{ g(x_n), x_1 - g_1(x_n), \ldots, x_n - g_n(x_n) \},
    \end{equation*}
    with $g, g_1,\ldots,g_n \in k[T]$, where $\operatorname{deg}(g) = D$ and $\operatorname{deg}(g_i) < D$ for $i=1,\ldots,n$.
\end{definition}

If the ideal $I$ is in shape position with respect to $x_n$, then the linear form $t := x_n$ and the polynomials $f := g$ and $f_i := -g_i f' \operatorname{mod} f$ for $i=1,\ldots,n$ yield a rational univariate representation of $V(I)$, where $f'$ denotes the derivative of $f$.
Note that $I$ is in shape position with respect to $x_n$ if and only if the linear form $t := x_n$ is injective on $V(I)$; in particular, every root is uniquely determined by its $x_n$-coordinate.

\subsection{From shape position to the general case}\label{sec:2-1}

As we have seen, ideals in shape position admit a particularly simple parametrization. In general, however, an ideal need not be in shape position with respect to any of its indeterminates.

Let $T$ be a new indeterminate and let
\[
t=a_1 x_1+\cdots+a_n x_n
\]
with $a_1,\ldots,a_n \in k$. Consider the ideal
\[
I_T := I+\langle T-t\rangle
~\subseteq~ k[T,x_1,\ldots,x_n],
\]
where $\langle T - t \rangle$ denotes the ideal generated by $T - t$. If $k$ is infinite, then there exists a choice of the coefficients $a_1,\ldots,a_n$ such that $\sqrt{I_T}$ is in shape position with respect to $T$~\cite{rouillier99}. 
In this case, and since the roots of $\sqrt{I_T}$ and $\sqrt{I}$ are in bijection, a rational univariate representation for $\sqrt{I_T}$ readily yields one for $\sqrt{I}$.

\subsection{Verifying the linear form}

One remaining task is to determine whether the chosen linear form $t=a_1x_1+\cdots+a_nx_n$ is injective on $V(I)$, or equivalently, whether $\sqrt{I_T}$ is in shape position with respect to $T$. Following~\cite{demin-rouillier-ruiz}, this question reduces to the bivariate elimination ideals:

\begin{lemma}
The ideal $I_T$ is in shape position with respect to $T$ if and only if,
for every $i=1,\ldots,n$, the ideal
$
I_T\cap k[T,x_i]
$
is in shape position with respect to $T$.
\end{lemma}

Thus, it suffices to verify the shape position property separately for each bivariate elimination ideal. For every $i=1,\ldots,n$, when the Gr{\"o}bner basis of
$
I_T\cap k[T,x_i]
$
in the lexicographic ordering with $T < x_i$ is known, one can apply the algorithm from~\cite{demin-rouillier-ruiz} to verify this. Moreover, for every $i=1,\ldots,n$, the polynomial $f_i(T)$ appearing in the parametrization can be recovered from the bivariate Gr{\"o}bner basis using the formulas from~\cite{demin-rouillier-ruiz}.

\section{Implemented algorithm}

Following the previous section, computing a rational univariate representation reduces to several computational tasks: computing the minimal polynomial of $T$ in $I_T \cap k[T]$, computing the lexicographic Gr{\"o}bner basis of bivariate elimination ideals $I_T \cap k[T,x_i]$ for $i=1,\ldots,n$, and verifying that the chosen linear form is injective. Once these are available, the polynomials $f_1,\ldots,f_n$ appearing in the parametrization are recovered directly from the bivariate Gr{\"o}bner bases.

We implement the algorithm due to Demin, Rouillier, and Ruiz~\cite{demin-rouillier-ruiz}, which can be seen as a variant of the FGLM algorithm~\cite{fglm}. The algorithm accepts as input the reduced degrevlex Gr\"obner basis of a zero-dimensional ideal $I$, and produces a parametrization of the roots of $\sqrt{I}$. It applies to arbitrary zero-dimensional ideals.

\begin{algorithm}
\caption{Parametrization of the roots of the radical}
\label{alg:rur}
\textbf{Input:}
A reduced Gr{\"o}bner basis $G$ of a zero-dimensional ideal $I \subseteq k[x_1,\ldots,x_n]$.

\textbf{Output:}
$f,f_1,\ldots,f_n \in k[T]$, a parametrization of the roots of $\sqrt{I}$.

\begin{enumerate}[label = \textbf{(\arabic*.)}, leftmargin=*, align=left, labelsep=\favoritelabelsep, itemsep=\favoriteitemsep]

\item \label{alg:step:mul-mat-1} Construct the multiplication matrices
$M_{x_1},\ldots,M_{x_n} \in k^{D\times D}$, as in~\Cref{sec:mul-tables-1}.

\item \label{alg:step:choose} Choose a linear form
$t = a_1 x_1+\cdots+a_nx_n$ with $a_1,\ldots,a_n \in k$.

\item \label{alg:step:mul-mat-2} Set $M_T := a_1 M_{x_1} + \ldots + a_n M_{x_n}$.

\item \label{alg:step:minpoly} For $i = 1,2,\ldots$ do \algorithmiccomment{{Compute the minimal polynomial}}
\begin{enumerate}
\item Compute $v(T^i) := M_{T} \; v(T^{i-1})$.
\item If there exist $c_0, \ldots, c_{i-1} \in k$ with $
 c_0 v(T^0) + \ldots + c_{i-1} v(T^{i-1}) + v(T^i) = 0
$, then set $d := i$ and $f := T^{d} + c_{d-1} T^{d-1} \ldots + c_0$ and {\bf break}.
\end{enumerate}

\item \label{alg:step:biv} For $i=1,\ldots,n$ \algorithmiccomment{{Compute bivariate elimination bases}}
\begin{enumerate}
\item Compute $G_i \subseteq k[T, x_i]$, the reduced Gr{\"o}bner basis of $I_T \cap k[T,x_i]$ in the lexicographical ordering with $T < x_i$, as in~\Cref{sec:bivariate}.
\end{enumerate}

\item \label{alg:step:test} For $i=1,\ldots,n$ \algorithmiccomment{{Test shape lemma}}
\begin{enumerate}
\item Test if $\sqrt{I_T} \cap k[T,x_i]$ is in shape position with respect to $T$ using $G_i$ with the test from~\cite{demin-rouillier-ruiz}. If the test fails, {\bf go to} step~\ref{alg:step:choose}.
\end{enumerate}

\item For $i = 1,\ldots,n$ do \algorithmiccomment{{Recover parametrization of coordinates}}
\begin{enumerate}
    \item Recover $f_i$ from $G_i$ using the formulas from~\cite{demin-rouillier-ruiz}.
\end{enumerate}

\item {\bf Return}
$
f,f_1,\ldots,f_n.
$
\end{enumerate}
\end{algorithm}

The algorithm is summarized in~\Cref{alg:rur}. For every monomial $m$, let $v(m) \in k^D$ denote the representation of $m$ in the quotient $k[x_1,\ldots,x_n]/I$ in the monomial basis $\mathcal{B}$. For every $i=1,\ldots,n$, denote by $M_{x_i}$ the matrix of the map 
\[\begin{aligned}M_{x_i} :~~ &k[x_1,\ldots,x_n]/I &\to &&k[x_1,\ldots,x_n]/I\\
&m &\mapsto &&x_i m.\end{aligned}\]

The algorithm first constructs the multiplication matrices $M_{x_1},\ldots,M_{x_n}$ and chooses a linear form $t$. Note that multiplication matrices for $I$ and $I_T$ coincide except for matrix $M_T$ for the newly introduced indeterminate $T$.
The minimal polynomial of $T$ in $I_T \cap k[T]$ is then obtained by finding a linear relation among the vectors $v(1),v(T),v(T^2),\ldots$. Let $d$ be the degree of the minimal polynomial.

Then, for every $i=1,\ldots,n$, the algorithm computes the reduced Gr{\"o}bner basis of the bivariate elimination ideal $I_T \cap k[T,x_i]$. This can be done by applying a part of the FGLM algorithm; for details, we refer to~\Cref{sec:bivariate}.
The bivariate bases are subsequently used both to verify that the chosen linear form places the radical in shape position and to recover the coordinate polynomials. The next section describes the implementation of these stages, with particular emphasis on the linear algebra.

\begin{remark}[On the choice of the linear form]
    Choosing the coefficients $a_1,\ldots,a_n$ at random in step~\ref{alg:step:choose} is a reliable way to obtain an injective linear form. Nevertheless, the algorithm may start with any linear form because injectivity is verified in step~\ref{alg:step:test}. In practice, one may first try sparse linear forms, for example with only a few nonzero coefficients.
\end{remark}

\section{Implementation details}\label{sec:impl}

\subsection{Multiplication matrices}\label{sec:mul-tables-1}

The first stage of~\Cref{alg:rur} constructs the multiplication matrices $M_{x_1},\ldots,M_{x_n}$ and $M_T$. This amounts to computing the columns of $M_{x_1},\ldots,M_{x_n}$.

Recall $G$ is the reduced degrevlex Gr{\"o}bner basis of $I$ and $\mathcal{B}$ is the monomial basis of $k[x_1,\ldots,x_n]/I$. Note the quotient algebras
$k[x_1,\ldots,x_n]/I$ and
$k[T,x_1,\ldots,x_n]/I_T$
share the monomial basis $\mathcal B$.
For every $w_j \in \mathcal{B}$ and every $x_i$ with $i=1,\ldots,n$, consider the product $m := w_j x_i$. 
Recall $v(m) \in k^D$ denotes the representation of $m$ in the quotient algebra.
Three cases may occur:
{\renewcommand{\labelenumi}{\textbf{Case \arabic{enumi}.}}
\begin{enumerate}
\item \label{case:basis} $m \in \mathcal{B}$; in this case, $v(m)$ is a standard unit vector.
\item \label{case:lt} $m \in \operatorname{lt}(G)$; in this case, $v(m)$ can be read off an element of $G$.
\item \label{case:other} $m \notin \mathcal{B} \cup \operatorname{lt}(G)$. We choose a variable $x_k$ dividing $w_j$ such that
\[
m' := m / x_k = (w_j / x_k) x_i
\]
Since $\mathcal{B}$ is closed under division, $(w_j/x_k) \in \mathcal{B}$, so $m'$ is again of the form $w x_i$ with $w \in \mathcal{B}$.
If $v(m') = a_1 w_1 + \cdots + a_D w_D$ for $a_1,\ldots,a_D \in k$, then
\begin{equation}\label{eq:mul-repr}
v(m) = v(m' x_k) = a_1 v(w_1 x_k) + \ldots + a_D v(w_D x_k).
\end{equation}
Note that $m' \notin \mathcal{B}$, otherwise~\eqref{eq:mul-repr} would be circular.
\end{enumerate}
}

To evaluate~\eqref{eq:mul-repr} for a particular $m$, we must know $v(w_i x_k)$ for every $a_i \neq 0$.
Hence, we traverse the monomials and compute $v(m)$ in an order that ensures that all vectors on the right-hand side of~\eqref{eq:mul-repr} have already been computed.

Once the vectors $v(w_j x_i)$ are computed, the matrices are known column-wise. In the implementation, columns corresponding to case~\ref{case:basis} are stored as unit vectors, whereas the remaining columns are stored densely.
Then, $v(1)$ is a unit vector, and $v(T^i) = M_{T} v(T^{i-1})$ for $i=1,2,\ldots$. We compute these matrix-vector products in step~\ref{alg:step:minpoly} of~\Cref{alg:rur} as a linear combination of the matrix columns. 

\subsection{Minimal polynomial}\label{sec:minpoly}

The second stage of~\Cref{alg:rur} computes the generator of the ideal $I_T \cap k[T]$, namely the minimal polynomial of $T$. This amounts to finding the first linear dependence over $k$ among $v(1)$, $v(T)$, $v(T^2)$, $\ldots$.

We construct the row echelon form incrementally. For every $i=1,2,\ldots$, let $V^{(i)} \in k^{i\times D}$ denote the matrix with the rows $v(1),\ldots,v(T^{i-1})$, which are obtained as in~\Cref{sec:mul-tables-1}.
For every $i=1,2,\ldots$, our implementation maintains and updates a decomposition
\begin{equation}\label{eq:kind-of-LU-decomposition}
L^{(i)} V^{(i)} = U^{(i)},
\end{equation}
where $L^{(i)} \in k^{i\times i}$ is lower triangular and $U^{(i)} \in k^{i\times D}$ is in row-echelon form.

\begin{algorithm}[H]
\caption{Update in incremental minimal polynomial computation}
\label{alg:minpoly-update}

\textbf{Input:} $V^{(i)}$, $L^{(i)}$, and $U^{(i)}$ as in
\eqref{eq:kind-of-LU-decomposition}, the vector $v(T^i)$.

\textbf{Output:}
either $c_0, \ldots, c_{i-1} \in k$, not all zero, such that $c_0v(T^0)+\cdots+c_{i-1}v(T^{i-1})+v(T^i)=0$, if such exist; otherwise, the updated matrices
$V^{(i+1)}$, $L^{(i+1)}$, and $U^{(i+1)}$.

\begin{enumerate}[label=\textbf{(\arabic*.)},
leftmargin=*,
align=left,
labelsep=\favoritelabelsep,
itemsep=\favoriteitemsep]

\item Set
$
r:=v(T^i)$ and $\gamma:=e_{i+1}
$, the standard unit vector.

\item Reduce $r$ with respect to the rows of $U^{(i)}$,
applying the same row operations on $\gamma$ using the corresponding rows of $L^{(i)}$.

\item If $r=0$, {\bf return} the relation $\gamma=(c_0,\ldots,c_{i-1},1)$.

\item Make $r$ monic and append it as a new row to $U^{(i)}$. Append the corresponding coefficient vector to $L^{(i)}$ and append $v(T^i)$ to $V^{(i)}$.

\item {\bf Return} $
V^{(i+1)},
L^{(i+1)},
U^{(i+1)}.$

\end{enumerate}
\end{algorithm}

At the first step, $v(1)$ is a unit vector, and we have
\[
V^{(1)} = [v(1)],
\qquad
L^{(1)} = [1],
\qquad
U^{(1)} = [v(1)],
\]
\Cref{alg:minpoly-update} is applied for $i=1,2,\ldots$ until a dependence is found. This occurs when $i=d$, where $d$ denotes the degree of the minimal polynomial.

In our implementation, since both $L^{(i)}$ and $U^{(i)}$ are triangular, we do not materialize them separately; instead, their entries are packed in a single dense matrix, so storing them requires storing only $d D$ field elements. 

We instantiate the algorithm over the finite field $k = \mathbb{F}_p$, where $p$ is a machine prime.
Row reduction in~\Cref{alg:minpoly-update} consists of repeated dense AXPY operations (updates of vectors of the form $y := \alpha x + y$).
With $p$ being a $28$-bit prime and using $64$-bit integers for representing field elements, we may accumulate $2^7$ such row multiples in one buffer before modular reduction is required. Overall, packing of the matrices $L^{(i)}$ and $U^{(i)}$ together with delaying modular reductions yields a simple inner loop that Julia automatically vectorizes using SIMD.

\subsection{Bivariate elimination bases}\label{sec:bivariate}

One of the final stages of~\Cref{alg:rur} computes, for every
$i=1,\ldots,n$, the reduced lexicographic Gr\"obner basis of the bivariate ideal $I_T\cap k[T,x_i]$.

Let $i \in \{1,\ldots,n\}$. Linear dependencies among the vectors $v(T^j x_i^k)$ with $j,k \in \mathbb{N}$ correspond to polynomial relations in $I_T \cap k[T,x_i]$. Consequently, constructing a row echelon form of these vectors yields a lexicographical Gr{\"o}bner basis of the ideal.

As in~\Cref{sec:minpoly}, we generate and reduce the vectors incrementally using Gaussian elimination. Whenever a newly generated vector is linearly dependent on the preceding ones, the corresponding dependency yields an element of the basis of $I_T\cap k[T,x_i]$.

The row echelon form of $v(1),v(T),\ldots,v(T^{d-1})$ has already been computed in~\Cref{sec:minpoly}. We therefore initialize the computation with this echelon basis and only generate the vectors $v(T^j x_i^k)$ involving positive powers of $x_i$, thus reusing the reductions of the pure powers of $T$.

\section{Motivation for dense linear algebra}\label{sec:motiv}

\subsection{On sparsity of multiplication matrices}\label{sec:on-sparsity}

The computation of the vectors $v(1), v(T), \ldots, v(T^d)$ is one of the main parts of~\Cref{alg:rur}. Sparse methods aim to exploit the sparsity of $M_T$ when repeatedly performing sparse matrix-vector products $v(T^i) := M_T v(T^{i-1})$ for $i=1,\ldots,d$. Consequently, the sparsity of $M_T$ is one of the main factors determining the practical efficiency of such methods.

{\setlength{\textfloatsep}{1pt plus 1pt minus 1pt}%
\setlength{\intextsep}{16pt plus 1pt minus 1pt}%
\begin{figure}[H]
\centering
\includegraphics[width=1.0\linewidth]{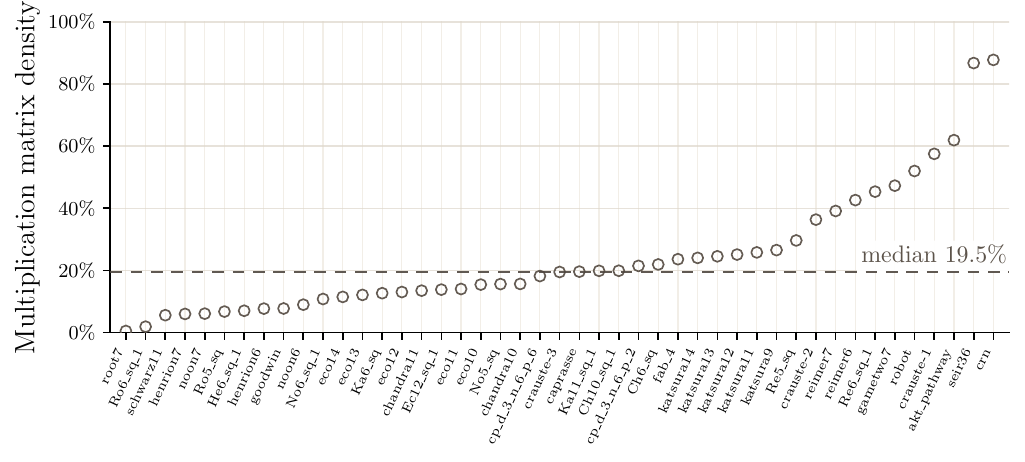}
\caption{Density of the multiplication matrix $M_T$, measured as the ratio of nonzero entries. In each example, the radical is put in shape position with respect to $T$. }
\label{fig:intro-density}
\end{figure}
}

Figure~\ref{fig:intro-density} reports the density of the multiplication matrix $M_T$ on our benchmark suite. We observe that the density is at least 10\% in $35$ out of $45$ examples, and the median density is approximately 19.5\%. Thus, the multiplication matrices encountered in practice can be far from sparse.

These observations suggest that sparse operations may not always be the most natural choice for computing $v(1), v(T), \ldots, v(T^d)$ in general. This motivates considering dense linear algebra as an alternative.

\begin{remark}[Shape position does not imply sparsity]
As we have seen, one may always construct $I_T \subseteq k[T,x_1,\ldots,x_n]$ whose radical is in shape position with respect to $T$. Recall 
$
I_T = I+\langle T-t\rangle,
$
where $t=a_1 x_1+\cdots+a_nx_n$ with $a_1,\ldots,a_n \in k$. The multiplication matrix of $T$ is then
\[
M_T=a_1 M_{x_1} + \cdots + a_n M_{x_n},
\]
which need not be sparse when several of $a_i$ are nonzero. Thus, although $\sqrt{I_T}$ is in shape position, the multiplication matrix of $T$ may be rather dense.
\end{remark}

\subsection{On the cost of Gaussian elimination}\label{sec:on-gauss-minpoly}

The computation of the coefficients $c_0,\ldots,c_{d-1}\in k$ in the relation $c_0 v(1) + c_1 v(T) + \ldots + v(T^d) = 0$ is the second major part of~\Cref{alg:rur}.

This relation can be computed by Gaussian elimination, as in~\Cref{sec:minpoly}. Alternatively, methods based on the Wiedemann algorithm replace Gaussian elimination by a Berlekamp--Massey reconstruction, whose cost is nearly linear in the degree of the minimal polynomial. However, they typically require at least twice as many matrix-vector multiplications, thus computing at least $2 d$ vectors instead of $d$. This speeds up the computation of the relation at the expense of additional matrix-vector products.

\begin{table}[H]
\centering
\setlength{\tabcolsep}{4pt}
\caption{The running time of two parts of~\Cref{alg:rur}: computing $v(1), \ldots, v(T^d)$ as in~\Cref{sec:mul-tables-1} and computing $c_0,\ldots,c_{d-1}$ via Gaussian elimination as in~\Cref{sec:minpoly}.}
\begin{tabular}{lrrr}
System & $D$ & Matrix-vector products & Gaussian elimination\\
\midrule
Chandra-13 & 4096 & 7.8 s & 7.8 s \\
Eco-14 & 4096 & 9.9 s & 7.8 s \\
Henrion-7 & 5040 & 4.1 s & 14.7 s \\
Noon-8 & 6545 & 10.8 s & 33.3 s\\
Katsura-14 & 8192 & 45.9 s & 71.0 s\\
Reimer-8 & 14400 & 101.7 s & 380.8 s\\
\end{tabular}
\label{table:pick-N}
\end{table}

Table~\ref{table:pick-N} reports the running time of these two tasks over $k := \mathbb{F}_p$, where $p$ is a machine prime. Our main observation is that, although Gaussian elimination is often more expensive, the cost of matrix-vector products is of the same order of magnitude. Thus, neither stage is a negligible part of the computation.

These measurements suggest that accelerating one part of the computation does not necessarily yield a much faster implementation. Moreover, if one performs matrix-vector products in the sparse regime in the Wiedemann algorithm, the computation of $2 d$ matrix-vector products instead of $d$ could itself become expensive when the matrix is not as sparse, e.g., as in examples in~\Cref{sec:on-sparsity}.

\subsection{Amortizing the cost of Gaussian elimination}

The previous subsection showed that Gaussian elimination is one of the computationally intensive parts of the algorithm. Nevertheless, this cost may be compensated by two additional benefits.

First, our goal is to compute a guaranteedly correct parametrization. The incremental Gaussian elimination in~\Cref{sec:minpoly,sec:bivariate} is fully deterministic. Consequently, in our implementation of step~\ref{alg:step:test} of~\Cref{alg:rur}, we are able to apply the test of~\cite{demin-rouillier-ruiz} to certify the computed parametrizations.

Second, as described in~\Cref{sec:bivariate}, the row echelon form computed during the minimal polynomial computation is reused in the construction of the bivariate elimination ideals. In particular, the reductions of the vectors $v(1)$, $v(T)$, $\ldots$, $v(T^{d-1})$
are performed only once and are reused throughout the remaining stages of~\Cref{alg:rur}. Thus, the cost of Gaussian elimination is amortized over the whole computation.

\section{Experimental results}

It is difficult to compare our implementation with msolve~\cite{msolve} or Giac~\cite{parisse-rur}.
Indeed, Giac can compute a parametrization only when the input ideal is radical. msolve can handle general ideals, but in our experiments it incurred a large overhead when the quotient is ``non-generic'' (following the terminology used in the msolve logs). When msolve encounters such non-generic quotient, it recomputes everything (the Gröbner basis and the parametrization), first attempting to change the monomial ordering and, if necessary, performing a generic change of coordinates.

On the other hand, our algorithm relies on the separation test~\cite{demin-rouillier-ruiz} to check if a linear form is injective and therefore requires less recomputation.
In both implementations, a reason for these repeated attempts is to test optimizations that are intended to improve efficiency in subsequent multi-modular computations for systems with rational coefficients, which is not in the scope of this paper.

\Cref{table:2} reports the performance of the implementations on benchmark systems that have been transformed into the radical shape lemma case by adding a separating linear form to the system.

\vspace{-1em}

\begin{table}
\centering
\setlength{\tabcolsep}{4pt}
\caption{The running time of msolve and our implementation (the timings include the computation of degrevlex Gr{\"o}bner bases).}
\begin{tabular}{lrrr}
System & $D$ & msolve & our method\\
\midrule
Katsura-13-EXT & 4096 & 68.0 s & 26.8 s\\
Chandra-13-EXT & 4096 & 54.0 s & 24.0 s \\
Eco-14-EXT & 4096 & 129.4 s & 39.9 s \\
Henrion-7-EXT & 5040 & 19.7 s & 20.5 s \\
Noon-8-EXT & 6545 & not generic & 37.5 s\\
No5-sq-a-EXT & 8192 & not generic & 91.4 s\\
Reimer-8-EXT & 14400 & 690.0 s & 516.8 s\\
\end{tabular}
\label{table:2}
\end{table}

To obtain a more representative view of the overall performance, it is also informative to consider the multi-modular computations on systems with rational coefficients (see \cite{demin-rouillier-ruiz}).

The experiments of \Cref{table:2} were performed on MacBook PRO MAX m2 running OSX 26.5 with 64 GB of RAM. Both implementations were run modulo the largest 28-bit prime. We used msolve~\cite{msolve} version~0.10.0, from the \href{https://msolve.lip6.fr/}{official website}, compiled with clang and the option \verb+-O3+. The binary was run with the default options. Our method uses the implementation in Julia in RationalUnivariateRepresentation.jl and uses Groebner.jl for degrevlex Gr{\"o}bner bases~\cite{gowda-demin}.

\section{Conclusion}

We presented a practical implementation for computing guaranteedly correct rational univariate representations of arbitrary zero-dimensional ideals based on the deterministic approach from~\cite{demin-rouillier-ruiz}. The implementation is available in the Julia package \href{https://newrur.gitlabpages.inria.fr/RationalUnivariateRepresentation.jl/}{RationalUnivariateRepresentation.jl}.

Our implementation is based on dense linear algebra and incremental Gaussian elimination. We argued that this design is well suited for practical computations: multiplication matrices are often only moderately sparse, the cost of matrix-vector products is often non-negligible, and the row echelon basis computed during the minimal polynomial computation is reused throughout the remaining stages of the algorithm. Experimental results show that these implementation choices lead to performance competitive with state-of-the-art software while producing certified parametrizations.

\bibliographystyle{splncs04}
\bibliography{refs}

@article{demin-rouillier-ruiz,
  author = {Demin, Alexander and Rouillier, Fabrice and Ruiz, Joao},
  title = {Reading Rational Univariate Representations on Lexicographic {G}r{"o}bner Bases},
  journal = {arXiv preprint arXiv:2402.07141},
  year = {2024},
  url = {https://arxiv.org/abs/2402.07141}
}

@ARTICLE{Wiedemann,
  author={Wiedemann, D.},
  journal={IEEE Transactions on Information Theory}, 
  title={Solving sparse linear equations over finite fields}, 
  year={1986},
  volume={32},
  number={1},
  pages={54-62},
  doi={10.1109/TIT.1986.1057137}}

@article{rouillier99,
  author = {Rouillier, Fabrice},
  title = {Solving Zero-Dimensional Systems through the Rational Univariate Representation},
  journal = {Applicable Algebra in Engineering, Communication and Computing},
  volume = {9},
  number = {5},
  pages = {433--461},
  year = {1999},
  doi = {10.1007/s002000050114}
}

@article{gowda-demin,
  author = {Gowda, Shashi and Demin, Alexander},
  title = {Groebner.jl: A Package for {G}r{"o}bner Bases Computations in Julia},
  journal = {arXiv preprint arXiv:2304.06935},
  year = {2023},
  url = {https://arxiv.org/abs/2304.06935}
}

@article{fglm,
  author = {Faug\`ere, Jean-Charles and Gianni, Patrizia and Lazard, Daniel and Mora, Teo},
  title = {Efficient Computation of Zero-Dimensional {G}r{\"o}bner Bases by Change of Ordering},
  journal = {Journal of Symbolic Computation},
  volume = {16},
  number = {4},
  pages = {329--344},
  year = {1993}
}

@article{faugere-mou,
  author = {Faug\`ere, Jean-Charles and Mou, Chenqi},
  title = {Sparse {FGLM} Algorithms},
  journal = {Journal of Symbolic Computation},
  volume = {80},
  pages = {538--569},
  year = {2017},
  doi = {10.1016/j.jsc.2016.07.025}
}

@article{hnrs-sparse-fglm,
  title = {Block-{K}rylov techniques in the context of sparse-{FGLM} algorithms},
  journal = {Journal of Symbolic Computation},
  volume = {98},
  pages = {163-191},
  year = {2020},
  note = {Special Issue on Symbolic and Algebraic Computation: ISSAC 2017},
  issn = {0747-7171},
  doi = {10.1016/j.jsc.2019.07.010},
  url = {https://www.sciencedirect.com/science/article/pii/S0747717119300756},
  author = {Hyun, Seung Gyu and Neiger, Vincent and Rahkooy, Hamid and Schost, {\'E}ric}
}

@inproceedings{msolve,
  author = {Berthomieu, J\'er\'emy and Eder, Christian and Safey El Din, Mohab},
  title = {\texttt{msolve}: A Library for Solving Polynomial Systems},
  booktitle = {Proceedings of the 2021 International Symposium on Symbolic and Algebraic Computation},
  year = {2021},
  pages = {51--58},
  doi = {10.1145/3452143.3465545}
}

@article{parisse-rur,
  author    = {Parisse, Bernard},
  title     = {Certifying a Probabilistic Parallel Modular Algorithm for Rational Univariate Representation},
  journal   = {arXiv preprint arXiv:2106.10912},
  year      = {2021},
  url       = {https://arxiv.org/abs/2106.10912}
}

\end{document}